\documentclass[aps,prl,twocolumn,superscriptaddress,groupedaddress]{revtex4-1}
\usepackage{graphicx,amssymb}
\usepackage{color,ulem}
\usepackage{bm}
 \usepackage{hyperref}

\newcommand{\eqref}[1]{(\ref{#1})}
\renewcommand{\Re}{\mathop \mathrm{Re}}
\renewcommand{\Im}{\mathop \mathrm{Im}}

\begin{document}
\title{
Supremacy of incoherent sudden cycles}

\author{Jukka P. Pekola}
\affiliation{QTF centre of excellence, Department of Applied Physics, Aalto University School of Science, P.O. Box 13500, 00076 Aalto, Finland}
\author{Bayan Karimi}
\affiliation{QTF centre of excellence, Department of Applied Physics, Aalto University School of Science, P.O. Box 13500, 00076 Aalto, Finland}
\author{George Thomas}
\affiliation{QTF centre of excellence, Department of Applied Physics, Aalto University School of Science, P.O. Box 13500, 00076 Aalto, Finland}
\author{Dmitri V. Averin}
\affiliation{Department of Physics and Astronomy, Stony Brook University, SUNY, Stony Brook, NY 11794-3800, USA}

\date{\today}

\date{\today}

\begin{abstract}
We investigate theoretically a refrigerator based on a two-level system (TLS) coupled alternately to two different heat baths. Modulation of the coupling is achieved by tuning the level spacing of the TLS. We find that the TLS, which avoids quantum coherences, creates finite cooling power for one of the baths in sudden cycles, i.e. acts as a refrigerator even in the limit of infinite operation frequency. By contrast, the cycles that create quantum coherence in the sudden expansions and compressions lead to heating of both the baths. We propose a driving method that avoids creating coherence and thus restores the cooling in this system. We also discuss a physical realization of the cycle based on a superconducting qubit coupled to dissipative LC-resonators.
\end{abstract}

\maketitle


{\sl Introduction:} In quantum thermodynamics, one of the timely questions is whether
and under what conditions quantum features such as entanglement and coherence can enhance the performance of heat engines and refrigerators \cite{alicki1979,Uzdin2015, Sai2016}. In many models of such machines, quantum
coherence is found to be useful \cite{Scully2003,Scully2011, Rahav2012,Jaramillo2016,hofer2016,Streltsov2017,Holubec2018,Camati}, whereas its adverse
effect has also been reported \cite{Kosloff2002,Thomas2014,Kilgour2018,Bayan2016,Brandner2016} or its usefulness may even depend on the quantity of interest \cite{Du2018}. An interesting regime is given by sudden cycles where control parameters of the system change infinitely rapidly. In this limit, the system poses potentially a powerful engine or refrigerator \cite{Paolo,Feldmann2000}. It has been suggested that refrigeration is made possible by quantum coherence in such cycles \cite{Uzdin2015,Feldmann2016,Newman}. Here we show in a simple yet realistic scheme that, on the contrary, an "incoherent" refrigerator which avoids creating off-diagonal elements of the density matrix in the eigenbasis of the instantaneous Hamiltonian, produces a finite cooling power in the sudden limit, while creation of coherence is a disadvantage and completely forbids cooling. Further, we demonstrate that it is possible to suppress coherence and thereby restore cooling in a quantum system in sudden cycles. For practical implementation, we present an experimentally feasible circuit using a superconducting qubit, where the presented cycle can be naturally realized. Our main results on the points above are captured by the final expressions in Eqs. \eqref{e11}, \eqref{e8} and \eqref{e8b}.

\begin{figure}[t!]
	\centering
	\includegraphics [width=0.9\columnwidth] {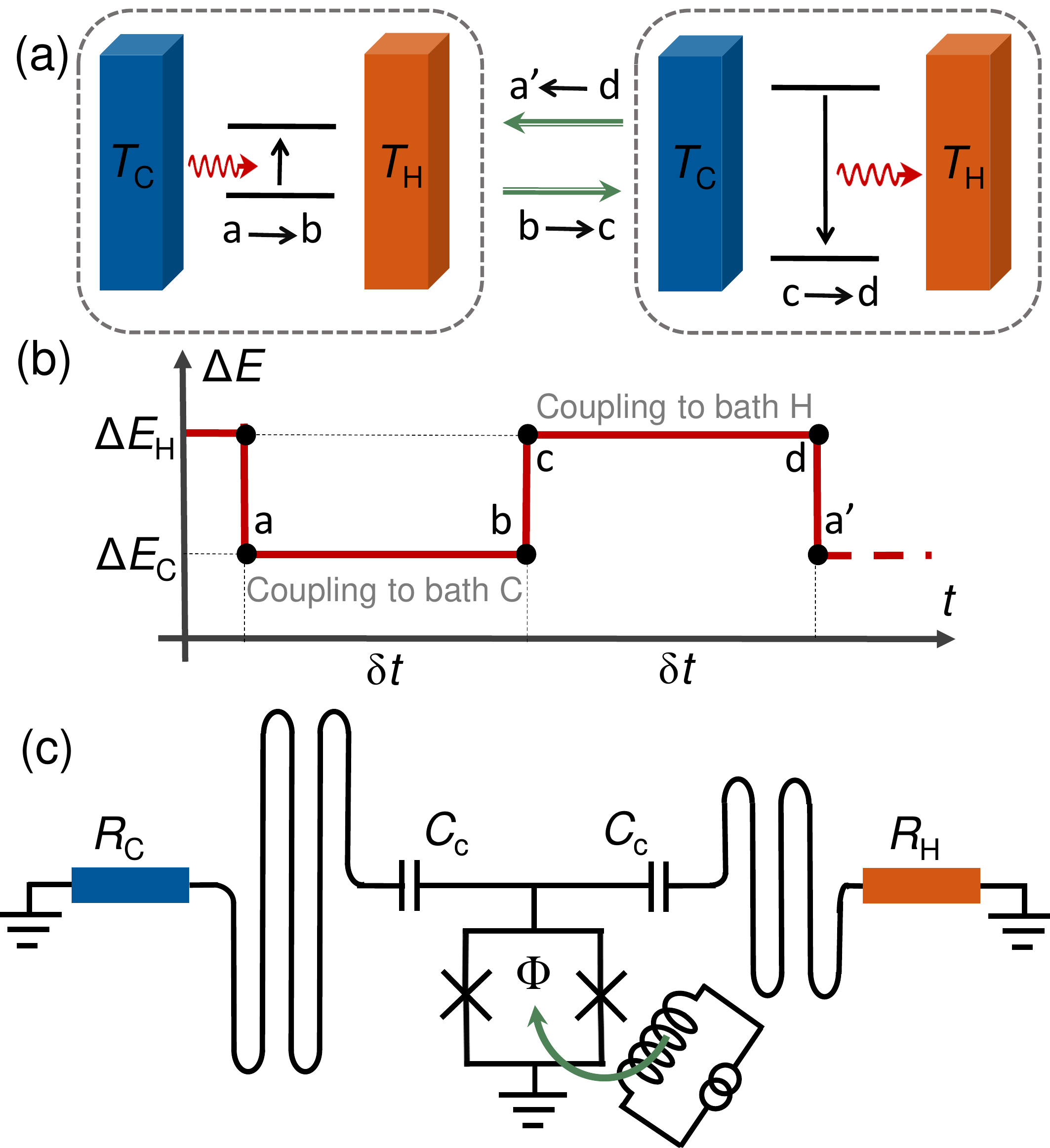}
	\caption{A two-level system (TLS) coupled to cold and hot baths at temperatures $T_{\rm C}$ and $T_{\rm H}$, respectively. (a) The cooling cycle where the TLS couples alternately to one of the baths at a time. The interaction of the TLS with each bath is controlled by the level separation: for small (large) splitting it exchanges energy with the cold (hot) bath. The green arrows depict the abrupt expansion (compression). (b) The driving protocol in time, demonstrating one cycle of the process in (a). In a sudden cycle $\delta t \rightarrow 0$. (c) Potential experimental realization: the schematic of a superconducting qubit capacitively ($C_{c}$) coupled to coplanar wave resonators, operating at two distinct frequencies, and terminated by resistors $R_{\rm C}$ and $R_{\rm H}$ acting as the heat baths. The energy separation $\Delta E$ of the TLS is tuned according to the protocol in (a) and (b) by applying magnetic flux $\Phi$.}
	\label{fig1}
\end{figure}

{\sl Description of the system:}
We first present an abstract model of our cooling cycle and then introduce a physical implementation of it based on a superconducting qubit. The idea is shown in Fig. \ref{fig1}a. A two-level system (TLS) is sandwiched between the two baths at temperatures $T_{\rm C}\equiv 1/(k_B\beta_{\rm C})$ and $T_{\rm H}\equiv 1/(k_B\beta_{\rm H})$. The essence of the cooling cycle is that when the level spacing is tuned to its low value $\Delta E_{\rm C}$, the system is coupled to the cold bath only, with the relaxation rate of $\Gamma_\downarrow ^{\rm C}$, and similarly, when the spacing assumes the higher value $\Delta E_{\rm H}$, the system couples only to the hot bath with $\Gamma_\downarrow^{\rm H}$. The excitation ($\uparrow$) and relaxation ($\downarrow$) rates induced by bath $\rm {B=C,H}$ satisfy the detailed balance condition,
\begin{equation} \label{db}
\Gamma_\uparrow^{\rm B}=e^{-\beta_{\rm B}\Delta E_{\rm B}}\Gamma_\downarrow^{\rm B}.
\end{equation}
Our suggested cooling cycle (Otto cycle) is as follows, see Fig. \ref{fig1}a and b: (i) The system is initially coupled to bath C with level spacing $\Delta E_{\rm C}$ for a time interval $\delta t$ (a $\rightarrow$ b). (ii) The level separation is increased ("{\sl abrupt compression}") from $\Delta E_{\rm C}$ to $\Delta E_{\rm H}$ (b $\rightarrow$ c). (iii) The system interacts with bath H at $\Delta E_{\rm H}$ for a short time interval $\delta t$ (c $\rightarrow$ d). (iv) The level separation is decreased ("{\sl abrupt expansion}") from $\Delta E_{\rm H}$ to $\Delta E_{\rm C}$ (d $\rightarrow$ a'). In the analysis below, we find a cyclic steady-state solution for the system state and heat currents, i.e. power, $P_{\rm C},P_{\rm H}$ to the cold and hot baths, respectively. In particular we look for the high frequency $f=1/(2\delta t)\rightarrow \infty$ solution under different conditions.

This cycle can be realized physically (Fig. \ref{fig1}c) by a superconducting qubit \cite{Koch2007,Clarke2008} coupled to the baths C and H via coplanar waveguide resonators with resonance frequencies $\omega_{\rm C}= \Delta E_{\rm C}/\hbar$ and $\omega_{\rm H}= \Delta E_{\rm H}/\hbar$, respectively  \cite{Bayan2016,Ronzani2018,Henrich2006}. If the difference between the level separations, $|\Delta E_{\rm H}-\Delta E_{\rm C}|$, is large enough for a given quality factor $Q_{\rm B}$ of the resonator, the TLS couples essentially to one bath only at a time and we obtain the presented alternating cycle as will be detailed in the final section of this paper.

{\sl Quantum cycle:}
The cycle described above can be analyzed precisely for a quantum TLS weakly coupled to the baths.
Due to piecewise constant and abrupt legs in the cycle, we do not have to resort to possibly uncontrolled master equations under rapid change of the parameters. Instead, we adopt the sudden approximation of quantum mechanics for the (de)compression legs, and the standard Lindbladian evolution \cite{breuer,jp2010} of the TLS with constant level separation for the thermalization legs over the time intervals $\delta t$. We analyze these two different types of evolutions one by one and then find the steady state (cyclic) result by imposing continuity of the density matrix in time.
The qubit has the Hamiltonian
\begin{equation}\label{i5}
H_{\rm Q}=-E_0(\Delta \sigma_x + q \sigma_z),
\end{equation}
where $E_0$ is the overall energy scale, $\Delta$ is the coupling and $q$ the control parameter (magnetic or electric field). Its eigenstates in the
computational basis $|+\rangle=(1~0)^{\dagger}$ and   $|-\rangle=(0~1)^{\dagger}$ read
\begin{eqnarray} \label{i6}
|g\rangle &=& 2^{-1/2}(\sqrt{1-\eta(q)}|-\rangle + \sqrt{1+\eta(q)}|+\rangle),\nonumber\\
|e\rangle &=&2^{-1/2}( \sqrt{1+\eta(q)}|-\rangle - \sqrt{1-\eta(q)}|+\rangle),
\label{q-dependence}
\end{eqnarray}
with level separation
\begin{equation}\label{i5}
\Delta E = 2E_0\sqrt{q^2+\Delta^2}.
\end{equation}
In Eq. \eqref{i6}, $\eta (q) \equiv (q/\Delta)/\sqrt{1+(q/\Delta)^2}$. We study the evolution  of the density matrix $\rho$ parametrized by  $\mathcal D \equiv \rho_{gg}-1/2$,  $\mathcal R\equiv \Re (\rho_{ge}e^{i\phi})$ and $\mathcal I \equiv \Im (\rho_{ge}e^{i\phi})$, where $\rho_{gg}=\langle g|\rho|g\rangle$, $\rho_{ge}=\langle g|\rho|e\rangle$, and $\phi=\int dt \Delta E /\hbar $ is the phase that could be accumulated along the thermalization legs. In our discussion below, we assume that this phase is not relevant, either because the overall operation cycle is so short that $\phi$ is negligible, or because the thermalization legs are timed so that $\phi$ is a multiple of $2\pi$. The latter regime can in principle be realized, since the thermalization legs that are short on the relaxation time scale, can be effectively of arbitrary length on the time scale set by the system energies. The relaxation can be neglected during the fast $q$-ramp between $q=0$ and $q=q_{\rm M}$, so that the density matrices $\rho^i,\rho^f$ before and after the ramp are connected by a unitary evolution, $\rho^f = U\rho^iU^\dagger$.  For a sudden ramp, $U=I$, i.e., $\rho^f=\rho^i$.
The eigenstates of the  initial and final
Hamiltonians of the ramp $q:0 \rightarrow q_{\rm M}$  is
obtained by substituting $q=0$ and $q=q_{\rm M}$, in Eq. (\ref{i6}), respectively.
In this ramp,  the elements of the final density matrix in the basis of the final Hamiltonian ($\{|g_f\rangle,|e_f\rangle\}$) can be written
as $\rho_{kl}^f \equiv \langle k|\rho^f |l\rangle =  \langle k|\rho^i |l\rangle=  \sum_{k'l'}\rho^i_{k'l'}\langle k|k'\rangle \langle l'|l\rangle$ where $|k\rangle(|l\rangle)$
denotes the eigenstates of the final Hamiltonian  and $|k'\rangle (|l'\rangle)$ represents the eigenstates of the  initial Hamiltonian.
Hence, during $q$-ramps: $0 \rightarrow q_{\rm M}$ ($b \rightarrow c$), the  elements of the final and the initial density
matrices in the  basis of their respective instantaneous Hamiltonians are related as:
\begin{eqnarray} \label{e6}
&&{\mathcal D}_{c} = \sqrt{1-\eta_{\rm M}^2}{\mathcal D}_{b}- \eta_{\rm M}{\mathcal R}_{b},\nonumber \\ &&{\mathcal R}_{c} = \sqrt{1-\eta_{\rm M}^2}{\mathcal R}_{b}+ \eta_{\rm M}{\mathcal D}_{b},
\end{eqnarray}
where $\eta_{\rm M} \equiv \eta(q_{\rm M})$. Similar analysis can also be made for the
ramp: $q_{\rm M} \rightarrow 0$ ($d\rightarrow a'$). For assumed real $\Delta$, the imaginary part $\mathcal I \equiv \Im (\rho_{ge}e^{i\phi})$ remains constant in these ramps.

For the (partial) thermalization parts of the cycle between the sudden legs, $q={\rm constant}$,  only the relaxation drives the TLS evolution i.e., according to the standard master equation we have
\begin{equation} \label{e8a}
\dot \rho_{gg}=-\Gamma_\Sigma^{\rm B} \rho_{gg} +\Gamma_\downarrow^{\rm B} ,\,\,\, \dot \rho_{ge} = -\frac{1}{2}\Gamma_\Sigma^{\rm B} \rho_{ge},
\end{equation}
where $\Gamma_\Sigma^{\rm B}\equiv \Gamma_\downarrow^{\rm B} +\Gamma_\uparrow^{\rm B}$.
In the limit of short thermalization time $\delta t$, we may then write with analogous notations as for the sudden leg,
\begin{eqnarray} \label{e9}
&&{\mathcal D}_f= {\mathcal D}_i+[\Gamma_\downarrow^{\rm B}-\Gamma_\Sigma^{\rm B} ({\mathcal D}_i+1/2)]\delta t \nonumber \\
&&{\mathcal R}_f = (1 -\frac{1}{2}\Gamma_\Sigma^{\rm B} \delta t){\mathcal R}_i,\,\,\,{\mathcal I}_f = (1 -\frac{1}{2}\Gamma_\Sigma^{\rm B} \delta t){\mathcal I}_i.
\end{eqnarray}
Equation \eqref{e9}, together with the fact that $\dot \mathcal I =0$ in the sudden legs, implies that ${\mathcal I} \equiv 0$ in a limit cycle.

Next, we combine all the four legs in the cycle assuming the steady-state situation when the system returns to the same state after each driving period (limit cycle, $\rho_{a'}=\rho_{a}$), obtaining a set of equations as 
\begin{eqnarray} \label{e10}
&&{\mathcal D}_b = {\mathcal D}_a+[\Gamma_\downarrow^{\rm C}-\Gamma_\Sigma^{\rm C} ({\mathcal D}_a+1/2)]\delta t ,\,\,{\mathcal R}_b = (1 -\frac{1}{2}\Gamma_\Sigma^{\rm C} \delta t){\mathcal R}_a \nonumber \\
&&{\mathcal D}_c = \sqrt{1-\eta_{\rm M}^2}{\mathcal D}_b -\eta_{\rm M}{\mathcal R}_b, \,\, {\mathcal R}_c
=  \sqrt{1-\eta_{\rm M}^2}{\mathcal R}_b+\eta_{\rm M}{\mathcal D}_b\nonumber \\
&&{\mathcal D}_d= {\mathcal D}_c+[\Gamma_\downarrow^{\rm H}-\Gamma_\Sigma^{\rm H} ({\mathcal D}_c+1/2)]\delta t ,\,\,{\mathcal R}_d = (1 -\frac{1}{2}\Gamma_\Sigma^{\rm H} \delta t){\mathcal R}_c\nonumber \\
&&{\mathcal D}_a =  \sqrt{1-\eta_{\rm M}^2}{\mathcal D}_d+ \eta_{\rm M}{\mathcal R}_d , \,\,{\mathcal R}_a =   \sqrt{1-\eta_{\rm M}^2}{\mathcal R}_d-\eta_{\rm M}{\mathcal D}_d.\nonumber \\
\end{eqnarray}
Heat currents to the cold $P_{\rm C}=\Delta E_{\rm C}({\mathcal D}_b-{\mathcal D}_a)/(2\delta t)$ and hot
$P_{\rm H}=\Delta E_{\rm H}({\mathcal D}_d-{\mathcal D}_c)/(2\delta t)$ baths are then given for $q_{\rm M}/\Delta\gg 1$ by
\begin{equation} \label{e11}
P_{\rm C(H)}=\Delta E_{\rm C(H)}\frac{\Gamma_\downarrow^{\rm C(H)}\Gamma_\Sigma^{\rm H(C)}(1-e^{-\beta_{\rm C(H)}\Delta E_{\rm C(H)}})}{4(2\Gamma_\Sigma^{\rm C(H)}+\Gamma_\Sigma^{\rm H(C)})}  > 0.
\end{equation}
Thus in this limit both baths are heated. As discussed below, this is a manifestation of the adverse effect of coherence on the performance of a quantum refrigerator. Based on Eq. (\ref{e10}), the heat currents to the cold and the hot baths can also be written for $\eta_{\rm M}\approx1$ as 
$P_{\rm C}=\Delta E_{\rm C}({\mathcal R}_c-{\mathcal R}_d)/(2\delta t)=\Delta E_{\rm C}\Gamma_\Sigma^{\rm H}{\mathcal R}_c/4$ and $P_{\rm H}=\Delta E_{\rm H}({\mathcal R}_b-{\mathcal R}_a)/(2\delta t)=-\Delta E_{\rm H}\Gamma_\Sigma^{\rm C}{\mathcal R}_a/4$,
showing an explicit relation between heat power and coherence. The lowest order correction to $P_{\rm C(H)}$ in $\Delta/q_{\rm M}$, $\delta P_{\rm C(H)}=\gamma_{\rm C(H)} \Delta/q_{\rm M}$, is obtained with
\begin{eqnarray} \label{e11b}
\gamma_{\rm C(H)}=-\Delta E_{\rm C(H)}\frac{(\Gamma_\downarrow^{\rm H(C)}-\Gamma_\uparrow^{\rm H(C)})\Gamma_\Sigma^{\rm C(H)}(\Gamma_\Sigma^{\rm C}+\Gamma_\Sigma^{\rm H})}{2 (2\Gamma_\Sigma^{\rm C}+\Gamma_\Sigma^{\rm H})(2\Gamma_\Sigma^{\rm H}+\Gamma_\Sigma^{\rm C})}. \nonumber
\end{eqnarray}

{\sl Classical Otto refrigerator at high frequency:}
For the classical system, we assume a diagonal density matrix whose evolution is governed by the rate equation for the ground state population $\rho_{gg}=1-\rho_{ee}$ as
\begin{equation} \label{e1}
\dot \rho_{gg}=\rho_{ee}\Gamma_{\downarrow}^{\rm B} -\rho_{gg} \Gamma_{\uparrow}^{\rm B} =\Gamma_{\downarrow}^{\rm B}-\Gamma_{\Sigma}^{\rm B} \rho_{gg}.
\end{equation}
Such dynamics can be realized for instance using a single-electron box as a classical TLS \cite{Dima}. For infinitely fast expansion and compression, $\rho$ again remains constant. Yet in the thermalization legs of infinitesimal duration, $\Gamma_{\Sigma}^{\rm B}\delta t\ll 1$, the population changes according to Eq. \eqref{e1} as
\begin{equation}\label{e2}
\rho_{gg}(\delta t)-\rho_{gg}(0)=[\Gamma_\downarrow^{\rm B}-\Gamma_\Sigma^{\rm B}\rho_{gg}(0)]\delta t.
\end{equation}
Here we have set the initial time in each thermalization leg to zero. In this situation the populations in the limit cycle are governed by
\begin{eqnarray}\label{e3}
&&{\mathcal D}_b={\mathcal D}_a+[\Gamma_\downarrow^{\rm C}-\Gamma_\Sigma^{\rm C}({\mathcal D}_a+1/2)]\delta t,\,\,{\mathcal D}_c={\mathcal D}_b,\nonumber\\&&{\mathcal D}_d={\mathcal D}_c+[\Gamma_\downarrow^{\rm H}-\Gamma_\Sigma^{\rm H}({\mathcal D}_c+1/2)]\delta t,\,\,{\mathcal D}_a={\mathcal D}_d ,
\end{eqnarray}
where ${\mathcal D}_i$ denotes the shifted ground state population $\rho_{gg}-1/2$, as before, at position $i=a,b,c,d$ in the cycle. 
We obtain in the linear order in $\delta t$,
$\Delta {\mathcal D}={\mathcal D}_b-{\mathcal D}_a={\mathcal D}_c-{\mathcal D}_d=(\Gamma_\downarrow^{\rm C}\Gamma_\uparrow^{\rm H}-\Gamma_\uparrow^{\rm C}\Gamma_\downarrow^{\rm H})\delta t/(\Gamma_\Sigma^{\rm C}+\Gamma_\Sigma^{\rm H}).$
From the detailed balance conditions \eqref{db}, the average power to the baths, $P_{\rm C(H)}=\pm \Delta {\mathcal D} \Delta E_{\rm C(H)} f$, is then
\begin{eqnarray}\label{e8}
P_{\rm C(H)}=&&\frac{1}{2}\frac{\Gamma_\downarrow^{\rm C}\Gamma_\downarrow^{\rm H}}{\Gamma_\Sigma^{\rm C}+\Gamma_\Sigma^{\rm H}}\\&&\times \big{(}e^{-\beta_{\rm H(C)}\Delta E_{\rm H(C)}}-e^{-\beta_{\rm C(H)}\Delta E_{\rm C(H)}}\big{)}\Delta E_{\rm C(H)}\nonumber.
\end{eqnarray}
One can see immediately from Eq. \eqref{e8} that for equal temperatures $\beta\equiv \beta_{\rm C}=\beta_{\rm H}$ and setting $\Delta E_{\rm H} > \Delta E_{\rm C}$,
\begin{equation} \label{e8b}
P_{\rm C} <0 ~~{\rm and}~~ P_{\rm H} >0,
\end{equation}
meaning that the bath to which the system couples at lower level splitting cools down whereas that with higher energy separation heats up. 
Equation (\ref{e8b}) is generally true for different temperatures when $\beta_{\rm H}\Delta E_{\rm H}>\beta_{\rm C}\Delta E_{\rm C}$. Thus incoherent dynamics leads to refrigeration even in sudden cycles. The coefficient of performance of the refrigerator is $\epsilon\equiv {\rm extracted~ heat/work}=-P_{\rm C}/(P_{\rm C}+P_{\rm H})$. Based on Eq. \eqref{e8}, in the sudden limit it is
\begin{equation}\label{e8c}
\epsilon=\frac{\Delta E_{\rm C}}{\Delta E_{\rm H}-\Delta E_{\rm C}},\nonumber
\end{equation}
which is precisely the same as for an ideal low frequency Otto cycle.

The expression of powers to the cold and hot baths can be further simplified if $\beta \Delta E\gg 1$ and assuming that the excitation when coupled to the cold bath presents the slowest rate. In this case $P_{\rm C}=-\Gamma_\uparrow^{\rm C}\Delta E_{\rm C}/2$.


The adverse effect of quantum coherence on refrigeration in our model can be further illustrated by the following considerations. Assume that a TLS starts from a state with the density matrix $\rho$ diagonal in the energy basis, and the occupation probability $P_{ee}$ of the excited state is smaller than the probability of the ground state. 
If it is then driven by a changing external control parameter, the final state of the system is $\rho'=U\rho U^{\dagger}$, where $U$ is the unitary evolution operator. Coherence can be created between the eigenstates of the final Hamiltonian if the system Hamiltonian does not commute at different time instances, $[H(t),H(t')]\neq0$. One can directly show that creation of the coherence in the final state, which depends upon the rate of driving, implies that the  occupation probability of the excited state $P_{ee}'=\langle e'|\rho'|e'\rangle$ at the end of the evolution ($|e'\rangle$ is the excited state of the final Hamiltonian) is higher than the initial $P_{ee}$. On the other hand, for an infinitely slow process, the quantum adiabatic theorem holds and hence no coherence will be created, i.e., the populations in the energy eigenstates remain unchanged. Therefore, in general, the final energy of a system which is driven fast is higher than that of a slowly driven system. This difference of energy can be interpreted as the cost of creating coherence. Further, if this system is allowed to interact with a heat bath, decoherence takes place, and the extra energy spent to create the coherence will be dissipated to the heat bath.
In quantum thermodynamics, this phenomenon is often called {\sl inner or intrinsic} friction \cite{Kosloff2002}. It has been studied in different contexts \cite{Thomas2014,Alecce2015,Deng2018}, and can be viewed as the reason for the failure of the quantum refrigerator in the high-frequency limit. 

The populations in the instantaneous eigenstates of
the Hamiltonian change under fast driving due to the creation of coherence. For example,
${\mathcal D}_c$ is different from ${\mathcal D}_b$ due to the sudden ramp in Eq. \eqref{e10}.
There are  'shortcut to adiabaticity'  protocols to keep the populations
unchanged during fast processes \cite{Chen2010,Bason2012,Deffner,Lutz,Zhang}.
The eigenvectors of the quantum TLS in  Eq. (\ref{q-dependence}) are $q$-dependent.
Therefore, when $q$ is varied in time, eigenvectors become time-dependent
which in turn creates coherence during the sudden cycle. As we have seen, creation of coherence affects refrigeration adversely. To suppress the creation of quantum coherence, we can use a simple and experimentally feasible technique \cite{Thomas2014}: we may envision a cycle, in which $q$ and $\Delta$ are varied in time such that their ratio remains constant throughout the cycle.
Since the energy eigenstates ($|g\rangle$, $|e\rangle$) in Eq. \eqref{i6} are functions of $q/\Delta$,
they become time-independent and hence no coherence will be created, but varying the parameters $q$ and $\Delta$ changes the energy level spacing (Eq. \eqref{i5}).
Since the density matrix in this protocol remains diagonal, the time evolution of the TLS
is governed by Eqs. \eqref{e1} and \eqref{e2} as in the classical regime. Therefore, the
refrigeration is restored and is described by the same Eqs. \eqref{e8}-\eqref{e8c} as above. The basic shortcuts to adiabaticity involve compensating fields that are proportional to the time derivative of $q$ \cite {Bason2012}. This means infinite fields for sudden cycles, which is infeasible for experimental realization. On the contrary, the protocol we propose above (constant $q/\Delta$) avoids this problem making it experimentally attractive. This can be realized for instance by tuning simultaneously magnetic flux and gate voltage in a charge qubit configuration \cite{Clarke2008,Makhlin}.

{\sl Experimental feasibility and discussion:}
Figure {\ref{fig1}c presents an experimental set-up proposed for realizing a four-stroke quantum refrigerator \cite{Bayan2016} that has been tested under steady-state conditions experimentally in Ref. \cite{Ronzani2018}.
In this circuit the alternating coupling between the two baths, resistors $R_{\rm C}$ and $R_{\rm H}$, is achieved thanks to the two $LC$ resonators with different frequencies $f_{\rm B}= \omega_{\rm B}/(2\pi)=\Delta E_{\rm B}/h$, $\rm{B=C,H}$. The rate of emitting a photon to bath B for a TLS with level separation $\Delta E$ is then obtained from the standard golden rule expression as \cite{Bayan2016}
\begin{eqnarray} \label{resrates}
\Gamma_\downarrow^{\rm B} = &&\kappa_{\rm B}\frac{\Delta^2}{q^2+\Delta^2}\frac{\Delta E/\hbar}{(1-e^{-\beta_{\rm B}\Delta E})}\nonumber\\&& \times{[1+Q_{\rm B}^2(\Delta E/\Delta E_{\rm B} -\Delta E_{\rm B}/\Delta E)^2]^{-1}}.
\end{eqnarray}
Here $\kappa_{\rm B}$ is the dimensionless coupling parameter between the qubit and the resonator, and $Q_{\rm B}$ is the quality factor of the lossy resonator B.
In Eq. (\ref{resrates}), the $q$-dependent coupling of noise is governed by $\Delta^2/(q^2+\Delta^2)$, the Lorentzian $Q_{\rm B}$ dependent denominator determines the $LC$-filtered bandpass of
the coupling, and $\Delta E/(1-e^{-\beta_{\rm B}\Delta E})$ is due to the bare thermal noise of the resistor. Thus, making the quality factor of the resonators $Q_{\rm B}$ much larger in comparison to $\Delta E_{\rm C}/(\Delta E_{\rm H}-\Delta E_{\rm C})$, the TLS couples essentially to one bath only at a time which helps us to ignore the possibilities of any unexpected behavior due to
different noise sources \cite{Affleck2003,Nalbach2017}. This condition can be met for any $Q_{\rm B} \gg 1$, unless the two resonators are nearly identical.
The regime we discuss, the "sudden limit", can be reached by operating at frequencies $f \gg \Gamma$, where $\Gamma$ can be approximated by Eq. \eqref{resrates} at resonance. This condition can be controlled by setting the coupling $\kappa_{\rm B}$ between the qubit and the resonator weak enough. Since this coupling is either capacitive or inductive in a superconducting qubit, it can be down-tuned by geometry of the device. Typical  numbers for superconducting (transmon) qubits are in the range of $\kappa_{\rm B}\sim 10^{-2}$ \cite{Koch2007,Ronzani2018}. For order of magnitude estimates, we may assume that the typical rates in Eq. \eqref{resrates} are $\Gamma \sim \kappa_{\rm B} \Delta E/\hbar$ at resonance. For a realistic level spacing of $\Delta E/k_{\rm B} = 0.1$ K, and $\kappa_{\rm B}$ cited above, we have $\Gamma \sim 100$ MHz; $f > 100$ MHz can be easily achieved in the experiment. In this situation, the typical powers, based on Eqs. \eqref{e11} and \eqref{e8} are of the order of $P_{\rm B} \sim \Gamma \Delta E \sim 10^{-16}$ W, which is about one to two orders higher than the experimental noise equivalent power achieved by standard bolometric techniques \cite{Ronzani2018}.
What is usually considered as the limit of validity of Markovian analysis, as presented here, is that the bath correlation time needs to be shorter than the inverse decay rates of the quantum system. This is achieved by down-tuning the qubit relaxation rates at resonance to below the typical electron-electron collision rate in metal absorbers and the inverse resonator linewidth $Q_{\rm B}/\omega_{\rm B}$, which both are $> 10^9$ Hz, corresponding to the relevant correlation time. It is to be noted that the equations for
the evolution of the density matrix we use are applicable to any equilibrium
reservoir regardless of its microscopic nature. In this respect, our model is
based on a fully realistic description of the heat baths.

In conclusion, we have demonstrated sudden cooling cycles for both classical and quantum systems. Quantum cycles lead to dissipation due to coherence generation. Yet refrigeration can be resumed by mimicing classical dynamics via a simple driving protocol, where the instantaneous eigenstates do not vary during the operation. Implementing the tunable coupling to the baths can turn out to be more challenging for a classical TLS \cite{Jonne}, since the diagonal evolution comes then at the cost of adding an uncontrollable decoherence path. Therefore, we propose that it is an advantage to use a quantum TLS avoiding coherences as explained. We present a realistic set-up based on superconducting circuit quantum electrodynamics platform to test our predictions experimentally.

{\it Acknowledgements:} We acknowledge Mikko M\"ott\"onen and Ken Funo for useful discussions. This work was funded through Academy of Finland grants 312057 and 303677 and from the European Union's Horizon 2020 research and innovation programme under the European Research Council (ERC) programme and Marie Sklodowska-Curie actions (grant agreements 742559 and 766025). G. T. thanks for the grant from the Centre for Quantum Engineering at Aalto University. D.V.A. is supported by the US NSF grant DMR-1836707.

\end{document}